\documentclass[prd,preprint,superscriptaddress,amsmath,amssymb,nofootinbib]{revtex4}
\usepackage{graphicx}
\usepackage{dcolumn}
\usepackage{bm}
\usepackage{amssymb}
\usepackage{amsmath}
\usepackage{epsfig}    
\usepackage{color}
\usepackage{slashed}
\usepackage{hhline}

\def\be{\begin{equation}}
\def\ee{\end{equation}}
\newcommand{\bea}{\begin{eqnarray}}
\newcommand{\eea}{\end{eqnarray}}
\newcommand{\nn}{\nonumber}

\begin{document}

{\begin{flushright}{KIAS-P20016, Pre2020 - 009}
\end{flushright}}

\title{A radiatively induced neutrino mass model with hidden local $U(1)$ and LFV processes $\ell_i \to \ell_j \gamma$, $\mu \to e Z'$ and $\mu e \to e e$ } 

\author{Takaaki Nomura}
\email{nomura@kias.re.kr}
\affiliation{School of Physics, KIAS, Seoul 02455, Korea}

\author{Hiroshi Okada}
\email{hiroshi.okada@apctp.org}
\affiliation{Asia Pacific Center for Theoretical Physics (APCTP) - Headquarters San 31, Hyoja-dong,
Nam-gu, Pohang 790-784, Korea}
\affiliation{Department of Physics, Pohang University of Science and Technology, Pohang 37673, Republic of Korea}

\author{Yuichi Uesaka}
\email{uesaka@ip.kyusan-u.ac.jp}
\affiliation{Faculty of Science and Engineering, Kyushu Sangyo University, 2-3-1 Matsukadai, Higashi-ku, Fukuoka 813-8503, Japan}

\date{\today}

\begin{abstract}

We investigate a model based on hidden $U(1)_X$ gauge symmetry in which neutrino mass is induced at one-loop level by effects of 
interactions among particles in hidden sector and the Standard Model leptons.
Neutrino mass generation is also associated with $U(1)_X$ breaking scale which is taken to be low to suppress neutrino mass. 
Then we formulate neutrino mass matrix, lepton flavor violating processes and muon $g-2$ which are induced via interactions among Standard Model leptons and particles in $U(1)_X$ hidden sector that can be sizable in our scenario.
Carrying our numerical analysis, we show expected ratios for these processes when generated neutrino mass matrix can fit the neutrino data.

\end{abstract}
\maketitle
\newpage

\section{Introduction}

A mechanism of generating non-zero neutrino masses is one of the important open questions in particle physics that requires extension of the standard model (SM).
In particular the tininess of neutrino masses would provide a hint for structure of physics beyond the SM.
In fact many mechanisms to generate tiny neutrino masses are discussed such as canonical (type-I) seesaw mechanism~\cite{Seesaw1, Seesaw2, Seesaw3, Seesaw4} 
 in which active neutrino mass is suppressed by heavy right-handed neutrino mass parameter.
Neutrino mass can be also suppressed when it is generated at loop level forbidding tree level generation~\cite{a-zee, Cheng-Li, Pilaftsis:1991ug, Ma:2006km,Babu:2002uu}; see also review paper ref.~\cite{Cai:2017jrq} and references therein. 
In such a case, particles in hidden sector often propagate inside a loop diagram generating neutrino mass.
Then a hidden $U(1)$ symmetry is one of the attractive candidates to control such a hidden sector forbidding tree level neutrino mass~\cite{Dey:2019cts, Nomura:2018kdi, Cai:2018upp, Nomura:2018ibs, Nomura:2017wxf, Ko:2017uyb, Ko:2016uft, Ko:2016wce, Ko:2016ala, Ko:2014loa, Ko:2014eqa,Ma:2013yga,Yu:2016lof,Ko:2016sxg}, where "hidden" $U(1)$ symmetry implies that SM particles do not have any charges under this symmetry and hidden sector indicates the field contents with hidden charges.

In a neutrino mass model with a hidden $U(1)$ symmetry  
radiatively generated Majorana neutrino mass is often associated with spontaneous breaking of such $U(1)$ symmetry.
For example, a loop diagram includes Majorana mass term of extra neutral fermion generated by a vacuum expectation value (VEV) of a scalar field which spontaneously breaks a hidden $U(1)$ gauge symmetry~\cite{Ko:2016sxg}.
In such a realization, small VEV has advantage of suppressing neutrino mass in addition to loop factor.  
Then tiny neutrino mass can be generated naturally and we would have sizable Yukawa interactions between hidden particles and SM leptons which are associated with neutrino mass generation.
Remarkably,  these interactions with sizable couplings can provide rich phenomenology such as lepton flavor violating (LFV) processes  $\ell_i \to \ell_j \gamma$, $\ell_i \to \ell_j \ell_k \bar \ell_l$ and $\mu e \to e e$.
Furthermore we would have light $Z'$ boson from hidden $U(1)$ breaking with small VEV and it can also provide LFV process such as $\mu \to e Z'$ at loop level. 
Investigation of such LFV ratios is important to test our scenario since these sizable Yukawa couplings are related to neutrino mass matrix and we would get information of flavor structure of the couplings by comparing various LFV ratios.  
We then study the LFV processes in one of the simplest scenario of one-loop radiative neutrino mass model with local hidden $U(1)$ symmetry. 

In this paper, we construct a neutrino mass model with hidden sector based on local $U(1)_X$ symmetry.
In our model, Majorana neutrino mass is generated at one-loop level where extra scalar boson and fermions propagate inside a loop.
Then neutrino masses are suppressed by loop factor and small mass difference between bosons from real and imaginary part of extra scalar field 
generated by a VEV breaking the local $U(1)_X$.
We formulate neutrino mass matrix, LFV processes and muon $g-2$ which are induced via interactions among SM leptons and particles in $U(1)_X$ hidden sector.
Then we perform numerical analysis searching for allowed parameter region and expected ratios for various LFV processes.

This paper is organized as follows.
In Sec.~II, we show our model and formulate 
neutrino mass generation mechanism,  LFVs  and  muon $g-2$ {in addition to scalar sector, $U(1)_X$ gauge sector, and  extra fermion sector}.
In Sec.~III, we carry out numerical analysis searching for allowed parameter sets and 
estimate ratios of LFV processes and muon $g-2$ with these parameters.
In Sec.~IV, we provide the summary of our results and the conclusion.

\section{Model}

\begin{table}[t!]
\begin{tabular}{|c||c|c||c|c||c|c|c|}\hline\hline  
& ~$L_L$~& ~$e_R$~  & ~$L'$~& ~$N'$~& ~$H$~ & ~$S$~& ~$\varphi$~ \\\hline
$SU(2)_L$ & $\bm{2}$  & $\bm{1}$  & $\bm{2}$  & $\bm{1}$  & $\bm{2}$  & $\bm{1}$   & $\bm{1}$    \\\hline 
$U(1)_Y$   & $-\frac12$ & $-1$ & $-\frac12$ & $0$  & $\frac12$  & $0$ & $0$  \\\hline
$U(1)_{X}$   & $0$ & $0$ & $Q_X$   & $Q_X$  & $0$  & $-Q_X$  & $2 Q_X$ \\\hline
\end{tabular}
\caption{ 
Charge assignments to fields in the model under $SU(2)_L\times U(1)_Y\times U(1)_X$ where we omitted quark sector since it is the same as the SM one.  }
\label{tab:1}
\end{table}

In this section, we introduce our model in which hidden local $U(1)_X$ symmetry is introduced.
As for new fermion sector, two kinds of vector fermions $L'$ and $N'$ with the same  $U(1)_X$ charge of $Q_X$, where $L'\equiv[N,E]^T$ is an isospin doublet and $N'$ is an isospin singlet. We assume these two fermions have three families.
Note that $U(1)_X$ gauge symmetry is anomaly free since the new fermions charged under it are vector-like.
As for scalar sector, we introduce SM singlet fields $S$ and $\varphi$ whose $U(1)_X$ charges are $-Q_X$ and $2 Q_X$ respectively, in addition to the SM-like Higgs field $H$. 
We summarize the charge assignments of the fields in Table~\ref{tab:1} where quark sector is omitted since it is the same as the SM.
Among the scalar fields, we require $H\equiv[h^+,(v+\tilde h+z)/\sqrt2]^T$ and $\varphi\equiv (v_\varphi+\varphi_R+i\varphi_I)/\sqrt2$ to develop VEVs while $S$ is an inert scalar field without a non-zero VEV.
Here, $h^+$ and $z$ are respectively absorb by the charged weak boson $W^+$ and the neutral one $Z$ in the SM,
while $\varphi_I$ is done by another neutral gauge boson $Z'$ in the hidden sector.
$U(1)_X$ is broken to remnant $Z_2$ symmetry where $L'$, $N'$ and $S\equiv (S_R+i S_I)/\sqrt2$ are odd while the other fields are even. 
Note that our model is one of the simplest local hidden $U(1)_X$ model realizing neutrino mass at one-loop level; another possibility is introducing inert doublet scalar instead of $L'$.

Under the symmetries in the model,  we write the relevant Yukawa interactions and Dirac mass term associated with extra fermions such that
\begin{align}
& -{\cal L_M}  =   M_{L'} \bar L' L' + M_{N'} \bar N' N',  \label{Eq:Mass} \\
& -{\cal L_\ell}
= y_{\ell} \bar L_L H e_R  +  f  \bar L_L  L'_R S  +  g \bar L'_L N'_R \tilde H + \tilde g \bar L'_R N'_L \tilde H + h_L \bar N'^c_L N'_L \varphi + h_R \bar N'^c_R N'_R \varphi
+ {\rm h.c.}, \label{Eq:yuk}
\end{align}
where $\tilde H = i \sigma_2 H^*$ $\sigma_2$ being second Pauli matrix, generation index is omitted, and $y_\ell$ can be diagonal matrix without loss of generality due to the redefinitions of the fermions.
The scalar potential is also given by
\begin{align}
V = & \mu_H^2 H^\dagger H + \mu_S^2 S^* S + \mu_\varphi^2 \varphi^* \varphi + \mu (S^2 \varphi + c.c.) \nonumber \\
&  + \lambda_H (H^\dagger H)^2 + \lambda_{\varphi} (\varphi^* \varphi)^2 + \lambda_S (S^* S)^2 \nonumber \\
& + \lambda_{H S} (H^\dagger H)(S^* S) + \lambda_{H \varphi} (H^\dagger H)(\varphi^* \varphi) + \lambda_{S \varphi} (S^*S)(\varphi^* \varphi), 
\end{align}
where we assume all couplings are real.

\subsection{Scalar sector}

In this subsection, we discuss mass spectrum in scalar sector of the model.
Firstly, we consider scalar bosons associated with $H$ and $\varphi$. 
The VEVs of the scalar fields, $v$ and $v_\varphi$, are derived by solving the stationary conditions 
$\partial V/ \partial v = \partial V/ \partial v_\varphi = 0$ such that
\begin{equation}
v = \sqrt{\frac{2 (\lambda_\varphi \mu_H^2 - \lambda_{H \varphi} \mu_\varphi^2 )}{\lambda_H \lambda_\varphi - \lambda_{H \varphi}^2 }}, \quad
v_\varphi = \sqrt{\frac{2 ( \lambda_H \mu_\varphi^2 - \lambda_{H \varphi} \mu_H^2) }{\lambda_H \lambda_\varphi - \lambda_{H \varphi}^2 }},
\end{equation}
where these values can be taken to be real positive without loss of generality.
We then obtain the squared mass terms for CP-even scalar bosons as
\begin{equation}
\mathcal{L} \supset \frac{1}{4} \begin{pmatrix} \tilde h \\ \varphi_R \end{pmatrix}^T \begin{pmatrix} \lambda_H v^2 & \lambda_{H \varphi} v v_\varphi \\  \lambda_{H \varphi} v v_\varphi  & \lambda_\varphi v_\varphi^2 \end{pmatrix} \begin{pmatrix} \tilde h \\ \varphi_R \end{pmatrix},
\end{equation} 
which can be diagonalized by an orthogonal matrix providing the mass eigenvalues of the form;
\begin{equation}
m_{h,h_D}^2 = \frac{\lambda_H v^2 +\lambda_\varphi v_\varphi^2 }{4} \pm \frac{1}{4} \sqrt{\left( \lambda_H v^2 -\lambda_\varphi v_\varphi^2 \right)^2 + 4 \lambda_{H \varphi}^2 v^2 v_\varphi^2 }.
\end{equation}
The corresponding mass eigenstates $h$ and $h_D$ are also given by   
\begin{equation}
\begin{pmatrix} h \\ h_D \end{pmatrix} = \begin{pmatrix} \cos \alpha & \sin \alpha \\ - \sin \alpha & \cos \alpha \end{pmatrix} \begin{pmatrix} \tilde h \\ \varphi_R \end{pmatrix}, \quad
\tan 2 \alpha = \frac{2 \lambda_{H \varphi} v v_\varphi}{\lambda_H v^2 - \lambda_\varphi v_\varphi^2},
\label{eq:scalar-mass-fields}
\end{equation}
where $\alpha$ is the mixing angle, and $h$ is identified as the SM-like Higgs boson.
In our scenario, the VEV of $\varphi$ is taken to be small as $\mathcal{O}(100)$ MeV for suppressing neutrino mass as we discuss below.
We also assume $\lambda_{H \varphi} \ll 1$ so that mixing angle $\alpha$ is negligibly small to avoid constraints from the SM Higgs measurements.
Thus $h$ is almost SM-like Higgs. The CP-odd components of $H$ and $\varphi$ are identified as Nambu-Goldstone bosons absorbed by $Z$ and $Z'$ bosons after symmetry breaking.
Note that we have remaining $Z_2$ symmetry after $U(1)_X$ symmetry breaking where $\{L', N', S \}$ are $Z_2$ odd, while the other fields including SM fields are $Z_2$ even due to the charge assignment. 

We next consider mass spectrum of inert scalar bosons from $S$.
The mass terms after symmetry breaking are given by
\begin{equation}
L_{m_S} =\frac{1}{2} \mu_S^2 (S_R^2 + S_I^2) + \frac{\mu v_\varphi}{ \sqrt{2}} (S_R^2 - S_I^2).
\end{equation}
Thus masses of $S_R$ and $S_I$ are 
\begin{align}
m_{S_R} = \sqrt{\mu_S^2 + \sqrt{2} \mu v_\varphi}, \\
m_{S_I} = \sqrt{\mu_S^2 - \sqrt{2} \mu v_\varphi},
\end{align}
where mass difference between real and imaginary part of $S$ is induced by coupling $\mu$.  
In our numerical analysis below, we parametrize the mass difference as $\Delta m_S \equiv m_{S_R} - m_{S_I} \propto \mu v_\varphi/\mu_S$.
We will take the mass difference to be as small as $\mathcal{O}(1)$ eV to $\mathcal{O}(100)$ eV, since $\mu$ is expected to be small as the corresponding operator breaks global symmetry~\footnote{Without $\mu$ term, the potential in Eq.(3) is invariant under the transformations; $H\to e^{ia}H$,$S\to e^{ib}S$, and $\varphi\to e^{ic}\varphi$
independently, where a, b, c are different number of integers. It means there are three global U(1) symmetries.
But, if $\mu$ term appears, we have a relation $2b+c\equiv0$(mod$2\pi$).
It suggests that one of the three U(1) symmetries is broken.
So, we are supposing this symmetry breaking $\mu$ term be tiny.
Another interpretation is to introduce Lepton number $L$, assigning $L(L_L)=L(e_R)=L(S)\neq0$ while the others zero.
Then, only the $\mu$ term violates the lepton number that can be interpreted that the small $\mu$ generates the tiny neutrino masses.
Furthermore, $L(\varphi)\neq0$ is possible to preserve the lepton number if $\varphi$ is an inert boson. It suggests that nonzero VEV of $\varphi$ also contributes to the neutrino masses. We would highly appreciate the referee to point  these fruitful interpretations out.
}
in the potential
and the scale of $v_\phi$ is also taken to be low. Such a tiny $\Delta m_S$ suppresses neutrino mass. 

\subsection{$Z'$ boson}
The Lagrangian for $U(1)$ gauge sector is given by
\begin{equation}
L_{G_{U(1)}} = -\frac{1}{4} B^{\mu \nu} B_{\mu \nu}  -\frac{1}{4} \tilde Z'^{\mu \nu} \tilde Z'_{\mu \nu} - \frac{1}{2} \sin \chi B^{\mu \nu} \tilde Z'_{\mu \nu},
\end{equation}
where $B^{\mu \nu}$ and $\tilde Z'^{\mu \nu}$ are field strength for $U(1)_Y$ and $U(1)_{X}$ gauge fields, and $\sin \chi$ is kinetic mixing parameter.
After $U(1)_X$ symmetry breaking by the VEV of $\varphi$, we obtain massive extra gauge boson $Z'$.
We derive $Z'$ mass such that 
\begin{equation}
m_{Z'} \simeq 2 Q_X g_X v_\varphi,
\end{equation}
where $g_X$ is the gauge coupling associated with $U(1)_X$ and $Z$--$Z'$ mixing effect is assumed to be small.
As we take $v_\varphi = \mathcal{O}(100)$ MeV, the mass of $Z'$ is $m_{Z'} \lesssim 100$ MeV in our scenario.
Notice here that the SM fermions generally interact with $Z'$ via mixing with the SM $Z$ boson in kinetic term. 
In fact the kinetic mixing is generated, even if we take $\sin \chi = 0$ at tree level, by $L'$ loop since it has both $U(1)_Y$ and $U(1)_X$ change.
We obtain at one loop level~\cite{Holdom:1985ag,Dienes:1996zr,Chiang:2013kqa} 
\begin{equation}
(\sin \chi)_{\text{one loop}} \simeq - \frac{g_X g_Y}{16 \pi^2} Q_X \ln \left( \frac{|q^2|}{M_{L'}^2} \right),
\end{equation}  
where $q$ is the renormalization scale and log-divergence is absorbed by bare kinetic mixing parameter.
Taking $q$ as electroweak scale, we find the one loop effect is $\mathcal{O}(10^{-3})$ and it can be safe from $Z$--$Z'$ mixing constraints~\cite{Langacker:2008yv}; $Z$--$Z'$ mixing angle is approximately given by $\sin \chi$.
Moreover, the size of mixing can be controlled with our convenient manner by tuning tree level contribution. 
Here, we assume that the mixing is tiny enough to neglect this mixing effect.

\subsection{Extra fermion sector}

In this subsection, we discuss mass spectrum in extra fermion sector.
The mass term of extra charged lepton is given by Dirac mass term of $L'$ as follows
\begin{equation}
M_{L'} \bar L' L' \supset M_{L'} \bar E E.
\end{equation}
In our model $E$ does not mix with SM charged leptons due to remnant $Z_2$ symmetry.

After symmetry breaking, mass terms of extra neutral fermions are obtained such that
\begin{align}
- L_{M_N} = & (M_{L'} \bar N_L N_R + M_{N'} \bar N'_L N'_R  + M_D \bar N_L N'_R + \tilde M_D \bar N_R N'_L + h.c.  )\nonumber \\
& + M_{N'_{LL}} \bar N'^c_L N'_L + M_{N'_{RR}} \bar N'^c_R N'_R,
\end{align}
where $M_D = g v /\sqrt{2}$, $\tilde M_D = \tilde g v /\sqrt{2}$ and $M_{N'_{LL(RR)}} = h_{L(R)} v_\varphi/\sqrt{2}$.
We then rewrite fields by $N_R \equiv X_1$, $N_L^c \equiv X_2$, $N'_R \equiv X_3$ and $N'^c_L \equiv X_4$, and 
Majorana mass matrix can be obtained as 
\begin{align}
L_{M_M} &=  \begin{pmatrix} \bar X^c_{1a} \\ \bar X^c_{2a} \\ \bar X^c_{3a} \\ \bar X^c_{4a} \end{pmatrix}^T 
\begin{pmatrix} 
{\bm 0}_{ab} & (M_{L'}^T)_{ab} & {\bm 0}_{ab} & (\tilde M_D^T)_{ab} \\ 
(M_{L'})_{ab} & {\bm 0}_{ab} &  (M_D)_{ab} & {\bm 0}_{ab} \\
{\bm 0}_{ab} & (M_D^T)_{ab} & (M_{N'_{LL}})_{ab} & (M_{N'}^T)_{ab} \\
 (\tilde M_D)_{ab} & {\bm 0}_{ab} &  (M_{N'})_{ab} & (M_{N'_{RR}})_{ab} 
\end{pmatrix}
\begin{pmatrix}  X_{1b} \\  X_{2b} \\  X_{3b} \\  X_{4b} \end{pmatrix} \nonumber \\
& \equiv \frac12 \bar X^c (M_X) X.
\end{align}
Then the mass matrix can be diagonalized by acting {a unitary}  
matrix 
as 
\begin{align}
V^T M_{X} V =D_N,  \quad X_{i a} =V_{ [a + 3 i -3] \alpha} \psi_\alpha^0 \label{eq:N-mix}
\end{align} 
where $\psi^0_\alpha$ is the mass eigenstate.

\subsection{Neutrino mass generation}

 \begin{figure}[tb]
\includegraphics[width=80mm]{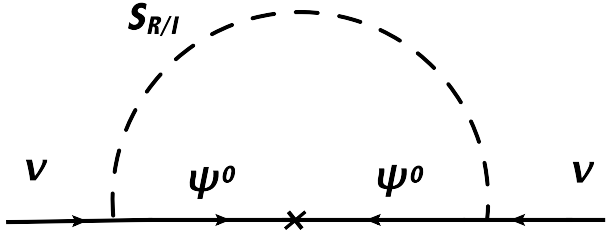}
\caption{One-loop diagrams generating neutrino mass.}
\label{fig:diagram-nu}
\end{figure}

In our model neutrino masses are generated via one-loop diagram shown in Fig.~\ref{fig:diagram-nu}.
Here we write the Yukawa interactions for neutrino mass generation in mass basis such that 
\begin{equation}
L \supset \frac{1}{\sqrt{2}} f_{i a} V_{a \alpha} \bar \nu_{L_i} \psi^0_\alpha (S_R + i S_I) + h.c.,
\end{equation}
where $V$ is the matrix diagonalizing extra neutral fermion matrix discussed above.
We then obtain neutrino mass matrix by calculating the diagram as 
\begin{align}
(m_\nu)_{ij} = & \sum_{a, b=1}^3 \sum_{\alpha=1}^{12}
 \frac{f_{ia} V_{a \alpha} (f_{jb} V_{b \alpha})^T}{32 \pi^2} 
M_{\psi^0_\alpha}  \left[ \frac{m^2_{S_R}}{m_{S_R}^2 - M^2_{\psi^0_\alpha} } \ln \left( \frac{m^2_{S_R} }{M^2_{\psi^0_\alpha}} \right) 
  -  \frac{m^2_{S_I}}{m_{S_I}^2 - M^2_{\psi^0_\alpha} } \ln \left( \frac{m^2_{S_I} }{M^2_{\psi^0_\alpha}} \right)  \right]\nn\\
  &=  f R f^T ,\\
  R&=  \frac{M_{\psi^0_\alpha} }{32 \pi^2} 
\left( V
 \left[ \frac{m^2_{S_R}}{m_{S_R}^2 - M^2_{\psi^0_\alpha} } \ln \left( \frac{m^2_{S_R} }{M^2_{\psi^0_\alpha}} \right) 
  -  \frac{m^2_{S_I}}{m_{S_I}^2 - M^2_{\psi^0_\alpha} } \ln \left( \frac{m^2_{S_I} }{M^2_{\psi^0_\alpha}} \right)  \right] V^T\right),
\end{align}
and the neutrino mass matrix is diagonalized by a unitary matrix $V_{MNS}$ as $D_\nu = V_{MNS}^T m_\nu V_{MNS}$.
Since $R$ is a symmetric matrix with three by three, Cholesky decomposition can be done as $R= T^T T$, where $T$ is an upper-right triangle matrix. 
$T$ is uniquely determined by $R$ except their signs, where we fix all the components of $T$ to be positive signs~\footnote{To see more concrete form of $T$, see ref.~\cite{Nomura:2016run} for example.}.
Then, the Yukawa coupling $f$ is rewritten in terms of the other parameters as follows~\cite{Casas:2001sr}:
\begin{align}
\label{Eq:Yukawa-result}
f&= V_{MNS}^* D_\nu^{1/2} V  {\cal O} (T^T)^{-1} ,
\end{align}
where ${\cal O}$ is three by three orthogonal matrix with an arbitrary parameters.
Note that $R$ is suppressed by $\Delta m_S$ and loop factor. Then Yukawa couplings $f_{ia}$ can have sizable values 
and significantly affect lepton flavor physics.
In our numerical analysis below, we impose $Max[|f|]\lesssim\sqrt{4\pi}$ as perturbative limit.

\subsection{$\ell_i \to \ell_j \gamma$ {and muon $g-2$}}

The relevant interaction to induce $\ell_i \to \ell_j \gamma$ lepton flavor violating(LFV) process is obtained from second term of Eq.~\eqref{Eq:yuk} as
\begin{equation}
\label{Eq:intLFV}
f_{i a} \bar L_L^i L'^a_R S + h.c. \supset f_{i a} \bar \ell_L^i E_R^a S +  f_{i a}^* \bar E_R^a \ell_L^i S^*.
\end{equation}
Considering one loop diagram, we obtain the BRs such that~\cite{Lindner:2016bgg}
\begin{align}
&{\rm BR}(\ell_i\to\ell_j\gamma)=
\frac{48\pi^3\alpha_{em}C_{ij}}{G_F^2 (4\pi)^4}
\left|\sum_{a} f_{j a} f^*_{i a} F(m_S,m_{E_a})\right|^2,\\
&F(m_a,m_b)\approx\frac{2 m^6_a+3m^4_am^2_b-6m^2_am^4_b+m^6_b+12m^4_am^2_b\ln\left(\frac{m_b}{m_a}\right)}{12(m^2_a-m^2_b)^4},
\end{align}
where $C_{21}=1$, $C_{31}=0.1784$, $C_{32}=0.1736$, $\alpha_{em}(m_Z)=1/128.9$, $G_F$ is the Fermi constant $G_F=1.166\times10^{-5}$ GeV$^{-2}$, and we have assumed to be $m_{\ell_{i,j}}^2/ m^2_{S,E_a}=0$ in our computation.
The current experimental upper bounds are given by~\cite{TheMEG:2016wtm, Aubert:2009ag,Renga:2018fpd}
\begin{align}
{\rm BR}(\mu\to e\gamma)\lesssim 4.2\times10^{-13},\quad 
{\rm BR}(\tau\to e\gamma)\lesssim 3.3\times10^{-8},\quad
{\rm BR}(\tau\to\mu\gamma)\lesssim 4.4\times10^{-8},
\label{eq:lfvs-cond}
\end{align}
where we impose these constraints in our numerical calculation.

In addition, we obtain contribution to muon $g-2$, $\Delta a_\mu$, through the same amplitude taking $\ell_i = \ell_j = \mu$ that approximately gives 
\begin{equation}
\Delta a_\mu \simeq \frac{m_\mu^2}{8 \pi^2} \sum_{a} f_{2 a} f^*_{2 a} F(m_S,m_{E_a}),
\end{equation}
where $m_\mu$ is the muon mass, and we have the same assumption as Eq.(22).
In our numerical analysis, we also estimate the value.

\subsection{Branching ratio of $\ell_i \to \ell_j \ell_k \bar \ell_l$}

\begin{figure}[tb]
\includegraphics[width=80mm]{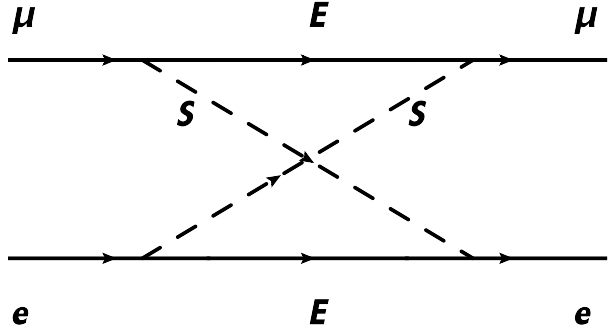}
\caption{The box diagram inducing $\ell_i \to \ell_j \ell_k \bar \ell_l$ decay and effective Lagrangian for $\mu e \to e e$ process.}
\label{fig:box-diagram}
\end{figure}

The LFV three body charged lepton decay processes are induced by box-diagram as shown in Fig.~\ref{fig:box-diagram}.
Calculating the one-loop diagram, we obtain BR for $\ell_i \to \ell_j \ell_k \bar \ell_l$ process such that 
\begin{align}
& {\rm BR}(\ell_i \to \ell_j \ell_k \bar \ell_l) \simeq \frac{m^5_{\ell_i} N_F}{6144 \pi^3 (4 \pi)^4 \Gamma_{\ell_i}} 
\left| \sum_{a,b=1}^{3} f_{ia} f^\dagger_{aj} f_{kb} f^\dagger_{b l} G(m_S, m_{E_a}, m_{E_b}) \right|^2, \\
& G(m_S, m_{E_a}, m_{E_b}) = \int_0^1 \frac{\delta(x+y+z-1) x}{x m_S^2 + y m^2_{E_a} + z m^2_{E_b}} dx dy dz,
\end{align}
where $\Gamma_{\ell_i}$ is the total decay width of $\ell_i$, 
$N_F =2$ for $\ell_i \to \ell_j \ell_j \bar \ell_j$ or $\ell_i \to \ell_k \ell_k \bar \ell_j$ and $N_F =1$ for $\ell_i \to \ell_j \ell_k \bar \ell_k$~\cite{Crivellin:2013hpa}.~\footnote{Although we have penguin types of diagrams at the one-loop level, we don't need to consider these constraints since we consider the constraints of $\ell_i\to \ell_j\gamma$.
This is because these penguin diagrams consist of the form of Eq.(24) and fine structure constant. Thus it is always small. See ref.~\cite{Toma:2013zsa} in details.}
{In our numerical analysis, we impose current experimental constraints~\cite{Bellgardt1988,Hayasaka2010}:
\begin{align}
& BR(\mu^+\to e^+e^+e^-)\lesssim 1.0\times 10^{-12}, \quad BR(\tau^\mp\to e^\pm e^\mp e^\mp) \lesssim2.7\times 10^{-8}, \nonumber \\
& BR(\tau^\mp\to e^\pm e^\mp\mu^\mp) \lesssim1.8\times 10^{-8}, \quad BR(\tau^\mp\to e^\pm\mu^\mp\mu^\mp) \lesssim1.7\times 10^{-8} \nonumber \\
& BR(\tau^\mp\to \mu^\pm e^\mp e^\mp) \lesssim1.5\times 10^{-8}, \quad BR(\tau^\mp\to \mu^\pm e^\mp\mu^\mp) \lesssim2.7\times 10^{-8} \nonumber \\
& BR(\tau^\mp\to \mu^\pm\mu^\mp\mu^\mp) \lesssim2.1\times 10^{-8}.
\end{align}
}

\subsection{$\mu \to e Z'$}

 \begin{figure}[tb]
\includegraphics[width=120mm]{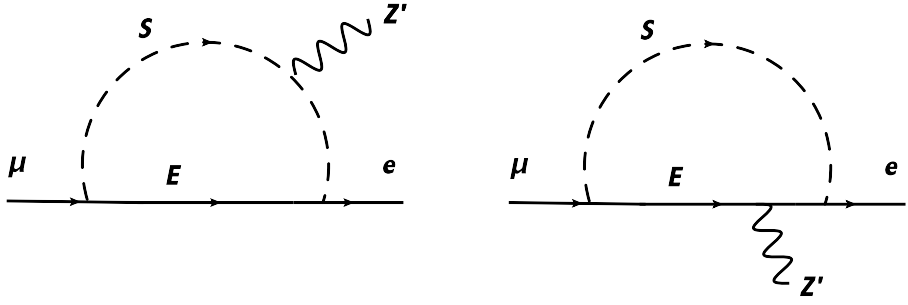}
\caption{One-loop diagrams inducing effective Lagrangian for $\mu \to e Z'$ process.}
\label{fig:diagram-meZp}
\end{figure}

In our scenario, $Z'$ is light and $\ell_i \to \ell_j Z'$ processes can be induced, where we focus on $\mu \to e Z'$ since it will be the clearest signal at experiments. 
Then its relevant interaction 
process arises from Eq.~\eqref{Eq:intLFV} with $U(1)_X$ gauge interaction of $E$ and $S$.
The $\mu \to e Z'$ process is obtained by one-loop diagrams as shown in Fig.~\ref{fig:diagram-meZp}.
Here we approximate as $m_{S_R} \simeq m_{S_I}$ and consider $S$ as a complex scalar boson in our calculation. 
Relevant effective Lagrangian is 
\begin{equation}
\mathcal{L}_{eff} = \frac{g_X}{2 \Lambda_L} (\bar e \sigma_{\alpha \beta} P_L \mu) X^{\alpha \beta} + \frac{g_X}{2 \Lambda_R} (\bar e \sigma_{\alpha \beta} P_R \mu) X^{\alpha \beta}, 
\end{equation}
where coefficients are estimated by calculating the diagrams.
We then obtain 
\begin{align}
 \frac{g_X}{\Lambda_{L(R)}} = & \left( \frac{g_X}{\Lambda_{L(R)}} \right)_{1} + \left( \frac{g_X}{\Lambda_{L(R)}} \right)_{2},  \nonumber \\
 \left( \frac{g_X}{\Lambda_{L}} \right)_{1} = & - \frac{m_e f_{1a} f^*_{2a} Q_X g_X}{8 \pi^2} \int [dX] \frac{zx}{\Delta_a},
 \quad \left( \frac{g_X}{\Lambda_{R}} \right)_{1} =  -\frac{m_\mu f_{1a} f^*_{2a} Q_X g_X}{8 \pi^2} \int [dX] \frac{yx}{\Delta_a}, \nonumber \\
  \left( \frac{g_X}{\Lambda_{L}} \right)_{2} = & -\frac{m_e f_{1a} f^*_{2a} Q_X g_X}{8 \pi^2} \int [dX] \frac{zx}{\Delta'_a},
 \quad \left( \frac{g_X}{\Lambda_{R}} \right)_{2} = - \frac{m_\mu f_{1a} f^*_{2a} Q_X g_X}{8 \pi^2} \int [dX] \frac{yx}{\Delta'_a}, \nonumber \\
\Delta_a = & -x(1-x) m^2_\mu - z(1-z)m^2_{Z'} - xz (m_\mu^2 + m^2_{Z'}) + (z+y) m^2_{E_a} + x m^2_S, \nonumber \\
\Delta'_a = & -y(1-y) m^2_\mu + (y+z) m^2_S  + y z (m_\mu^2 - m^2_{Z'}) + x m^2_{E_a}, 
\end{align}
where $\int d[X] = \int_0^1 dx dy dz \delta(1-x-y-z)$.
Note that $\Lambda_L/\Lambda_R = m_\mu/m_e$ in our mode and $\mu \to e Z'$ process is dominantly induced by $g_X/\Lambda_R$ effect.
In terms of the effective couplings, the branching ratio is given as
\begin{equation}
BR\left(\mu\to eZ'\right)=\frac{12\pi^2}{G_F^2m_\mu^2}\left(\left| \frac{g_X}{\Lambda_L} \right|^2 + \left| \frac{g_X}{\Lambda_R} \right|^2\right)\left(1-\frac{m_{Z'}^2}{m_\mu^2}\right)\left(1-\frac{m_{Z'}^2}{2m_\mu^2}-\frac{m_{Z'}^4}{2m_\mu^4}\right),
\end{equation}
In our numerical analysis below, we impose the constraint	
\begin{equation}
\label{Eq: MuEX-const}
\left| \frac{g_X}{\Lambda_L} \right|^2 + \left| \frac{g_X}{\Lambda_R} \right|^2 < \left( \frac{1.8 \times 10^{-10} }{\rm GeV} \right)^2.
\end{equation}
This bound is obtained from $BR(\mu \to e X) < 2.6 \times 10^{-6}$ with massless particle $X$ \cite{Jodidio1986}.

\subsection{$\mu e \to e e$}
In our model $\mu e \to e e $ process in a muonic atom \cite{Koike2010} is also induced by Eq.~\eqref{Eq:intLFV}.
We then obtain relevant effective interactions from the same diagram inducing $\mu \to e \gamma$ and the box-diagram shown in Fig.~\ref{fig:box-diagram} such that
\begin{equation}
\mathcal{L}_{\rm eff} = - \frac{4 G_F}{\sqrt{2}} m_\mu [A_R \bar e \sigma^{\alpha \beta} P_R \mu + A_L \bar e \sigma^{\alpha \beta} P_L \mu] F_{\alpha \beta}  -  \frac{4 G_F}{\sqrt{2}} g_4 [\bar e \gamma^\alpha P_L \mu][ \bar e \gamma_\alpha P_L e] + h.c. \, ,
\end{equation}
where the coefficients $A_R$, $A_L$, and $g_4$ in our model are derived as
\begin{align}
& A_R \simeq \frac{e}{16 \pi^2} \frac{\sqrt{2}}{4 G_F} \sum_{a} f_{1 a} f^*_{2 a} F(m_S,m_{E_a} ) , \\
& A_L \simeq \frac{e}{16 \pi^2} \frac{\sqrt{2}}{4 G_F} \frac{m_e}{m_\mu} \sum_{a} f_{1 a} f^*_{2 a} F(m_S,m_{E_a} ) , \\
& g_4 = \frac{\sqrt{2}}{4 G_F} \sum_{a,b} \frac{(f_{1a} f^*_{2 a}) (f_{1b} f^*_{1 b})}{32 \pi^2} G(m_S, m_{E_a}, m_{E_b}).
\end{align}
Here $A_L$ is suppressed by $m_e/m_\mu$ compared to $A_R$.
The ratio of $\mu e \to e e$ width and total decay width of muonic atom can be estimated by $A_{L,R}$ and $g_4$ where we denote the ratio by $R_{\mu e \to e e}$. 
The $R_{\mu^-e^-\to e^-e^-}$ is represented as
\begin{align}
R_{\mu^-e^-\to e^-e^-}=&\frac{\tilde{\tau}_\mu G_F^2}{\pi^3}\int_{m_e}^{m_\mu-B_\mu^{1s}-B_e^{1s}}dE_1\left|\bm{p}_1\right|\left|\bm{p}_2\right| \nonumber\\
&\times\sum_{\kappa_1,\kappa_2,J}\left(2J+1\right)\left(2j_{\kappa_1}+1\right)\left(2j_{\kappa_2}+1\right)\left|A_LW_L+A_RW_R+g_4W_4\right|^2,
\end{align}
where $\tilde{\tau}_\mu$ is the lifetime of a muonic atom, which is given in Ref.~\cite{Suzuki1987}.
$B_\ell^{1s}$ ($\ell=\mu,e$) is the binding energy of the initial lepton $\ell$ in a $1s$ state.
For simplicity, we take into account only the $1s$ electrons, which give the dominant contribution.
This formula of $R_{\mu^-e^-\to e^-e^-}$ includes the numerical integration by the energy $E_1$ of one emitted electron.
Once $E_1$ is fixed, the energy of the other emitted electron is determined by $E_2=m_\mu+m_e-B_\mu^{1s}-B_e^{1s}-E_1$ due to the energy conservation.
$J$ is the total angular momentum of the lepton system, and $\kappa_n$ ($n=1,2$) indicates the angular momentum of each electron.
The explicit formulas of $W_i$s ($i=L,R,4$) are given in Refs.~\cite{Uesaka2016,Uesaka2018}.

The $R_{\mu^-e^-\to e^-e^-}$ gets larger in a muonic atom with a larger proton number.
In our calculation, we assume the use of muonic lead ($^{208}$Pb).

\section{Numerical analysis}

In this section numerical analysis is carried out where we search for allowed values of free parameters satisfying neutrino data 
and show ratios for LFV processes as well as muon $g-2$ estimated by the allowed parameter sets.

We scan relevant free parameters in our model in the following region:
\begin{align}
& m_{S_R} \in [100, 1000] \ {\rm GeV}, \quad \Delta m_S \in [10^{-9}, 10^{-7}] \ {\rm GeV}, \quad g_X \in [0.1, 1.0], \nonumber \\
& \left\{(M_{L'})_{ab}, (M_{N'})_{ab} \right\} \in [100, 1000] \ {\rm GeV}, \quad  \left\{(M_D)_{ab}, (\tilde M_D)_{ab} \right\} \in [100, 500] \ {\rm GeV}, \nonumber \\
& \left\{ (M_{N'_{LL}})_{ab}, (M_{N'_{RR}})_{ab} \right\} \in [0.01, 1] \ {\rm GeV},
\end{align}
where we fix $v_\varphi = 100$ MeV and $Q_X = 1$. 
Note that the scales of mass matrix are chosen taking into account the fact that $\{M_{D}, \tilde M_{D} \} \propto v$ and $M_{N'_{LL(RR)}} \propto v_\varphi$ while $M_{L', N'}$ are bare mass parameters. 
Then we search for the allowed parameter sets which satisfies neutrino data of recent global fit by NuFIT 4.1~\cite{Esteban:2018azc,Nufit} 
\begin{align}
& | \Delta m^2_{\rm atm}|=[2.436-2.618]\times 10^{-3}\ {\rm eV}^2,  \quad  \Delta m^2_{\rm sol}=[6.79-8.01]\times 10^{-5}\ {\rm eV}^2,\nonumber \\
&\sin^2\theta_{13}=[0.02044-0.02435],  \quad  \sin^2\theta_{23}=[0.433-0.609],\nonumber \\
& \sin^2\theta_{12}=[0.275-0.350],
\end{align}
where we consider normal ordering (NO) case and Dirac(Majorana) CP phases are taken to be $[0, 2 \pi]$.
Then Yukawa couplings $f_{i \alpha}$ are determined by Casas-Ibarra parametrization in Eq.~\eqref{Eq:Yukawa-result} with $Max[|f|]\lesssim \sqrt{4\pi}$. 
Also, we impose the upper bound for the lightest neutrino mass to be 0.1 eV in our numerical analysis.

 \begin{figure}[tb]
\includegraphics[width=80mm]{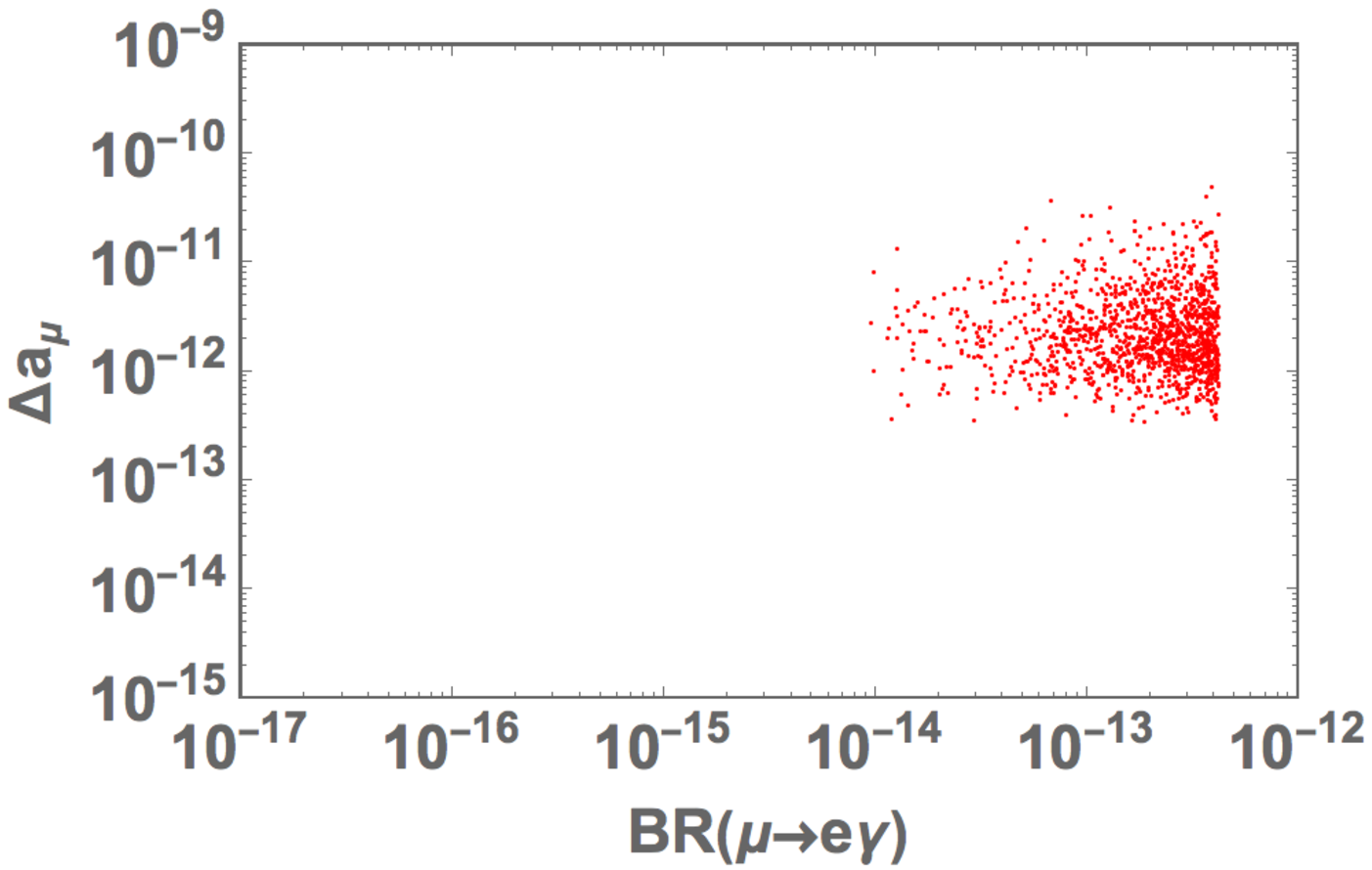} \
\includegraphics[width=80mm]{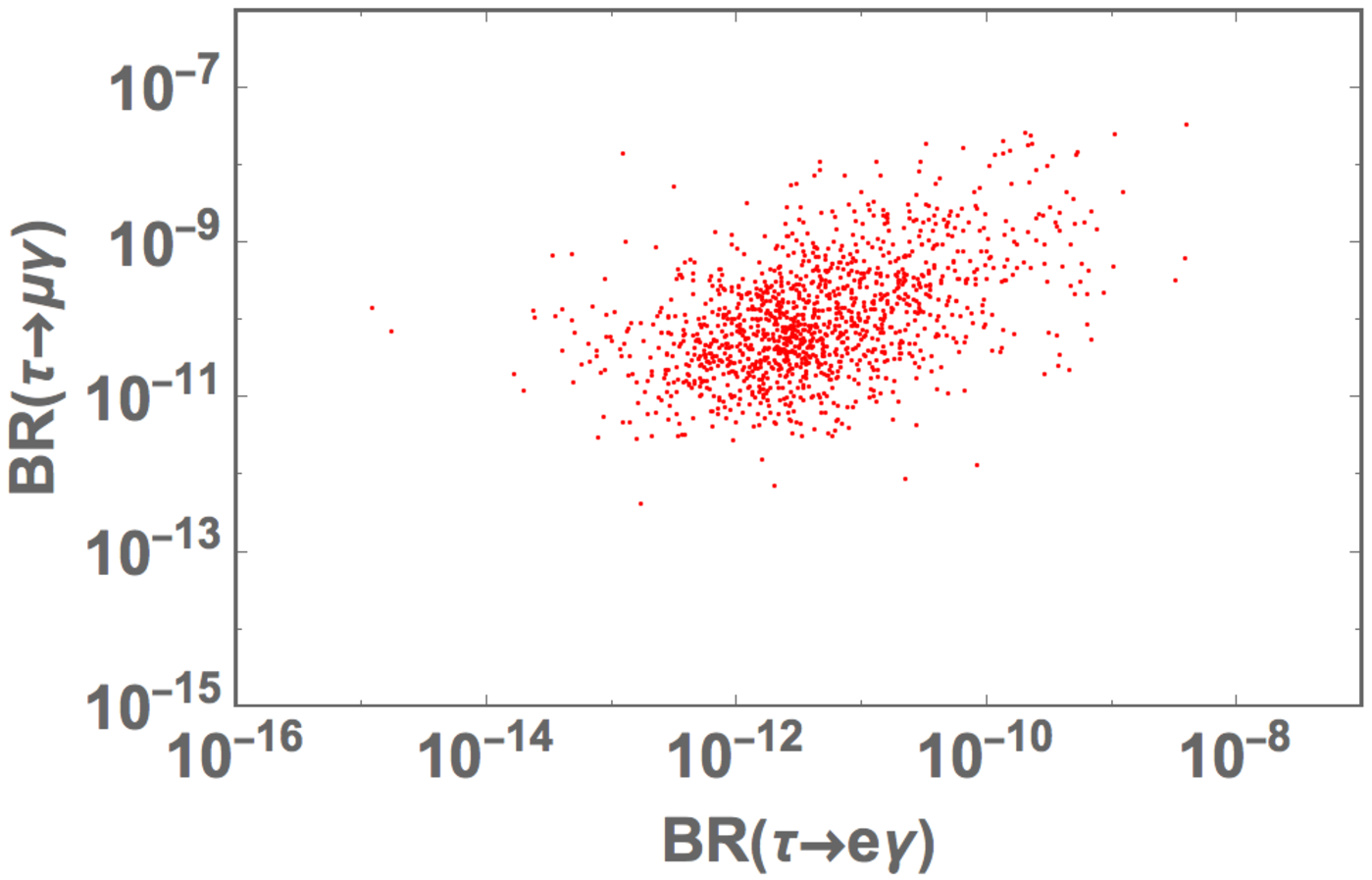}
\caption{BRs for $\ell_i \to \ell_j \gamma$ and $a_\mu$ for allowed parameter sets.}
\label{fig:LFV1}
\end{figure}

 \begin{figure}[tb]
\includegraphics[width=80mm]{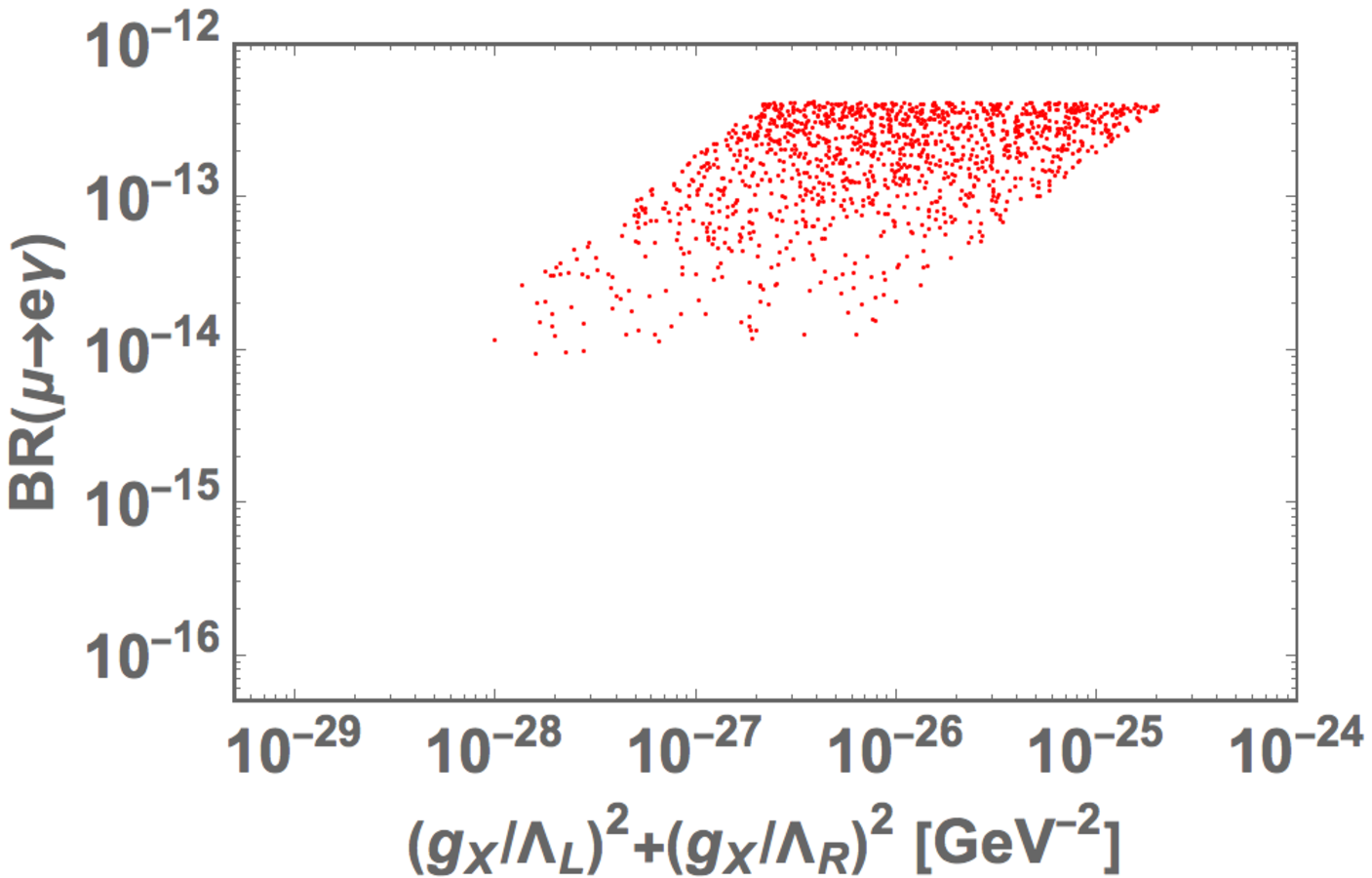}
\includegraphics[width=80mm]{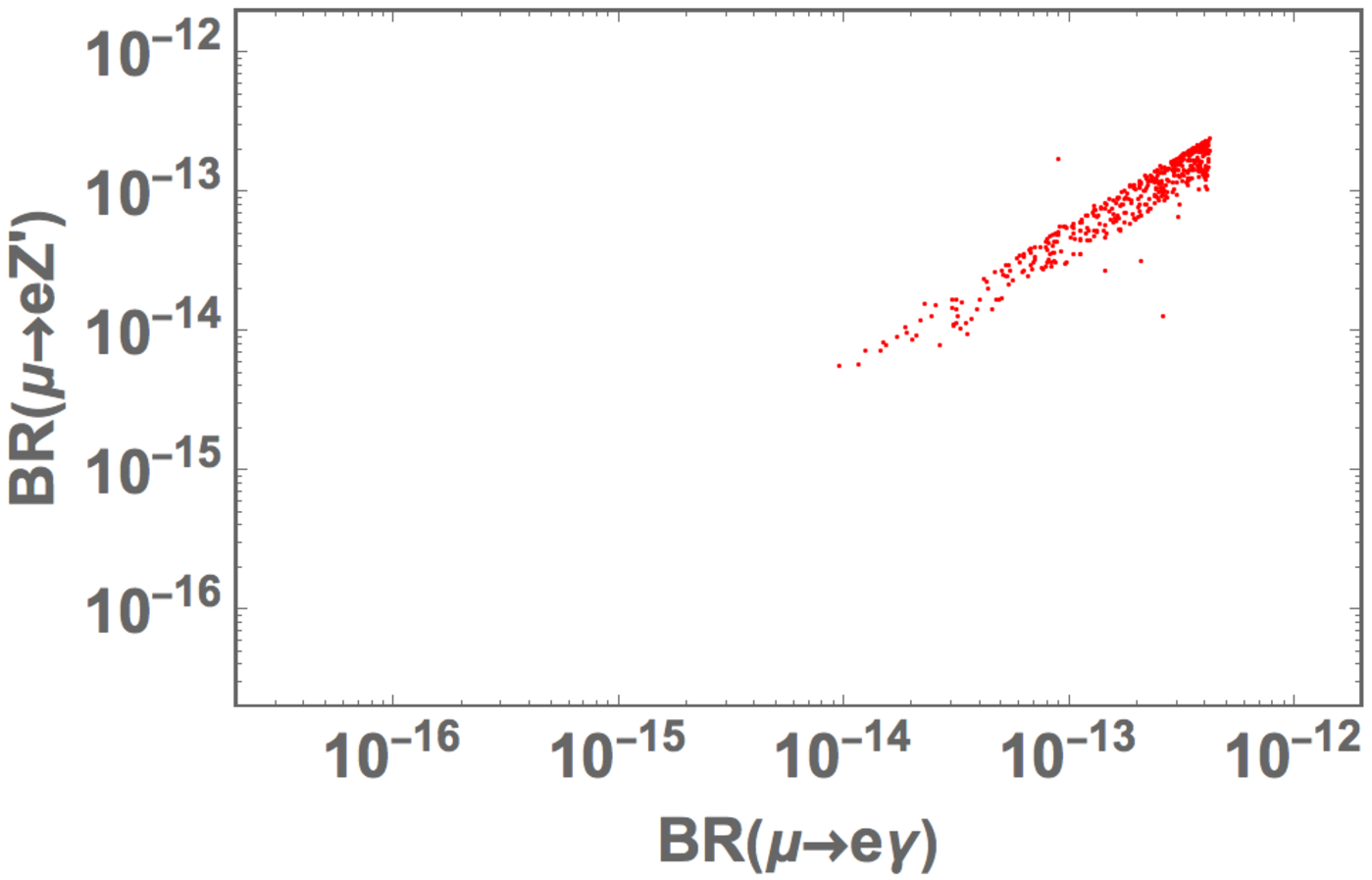}
\caption{Left: $BR(\mu \to e \gamma)$ and $(g_X/\Lambda_L)^2 +(g_X/\Lambda_R)^2$ for allowed parameter sets. Right: $BR(\mu \to e \gamma)$ and $BR(\mu \to e Z')$ for allowed parameter sets.}
\label{fig:LFV2}
\end{figure}

 \begin{figure}[tb]
\includegraphics[width=80mm]{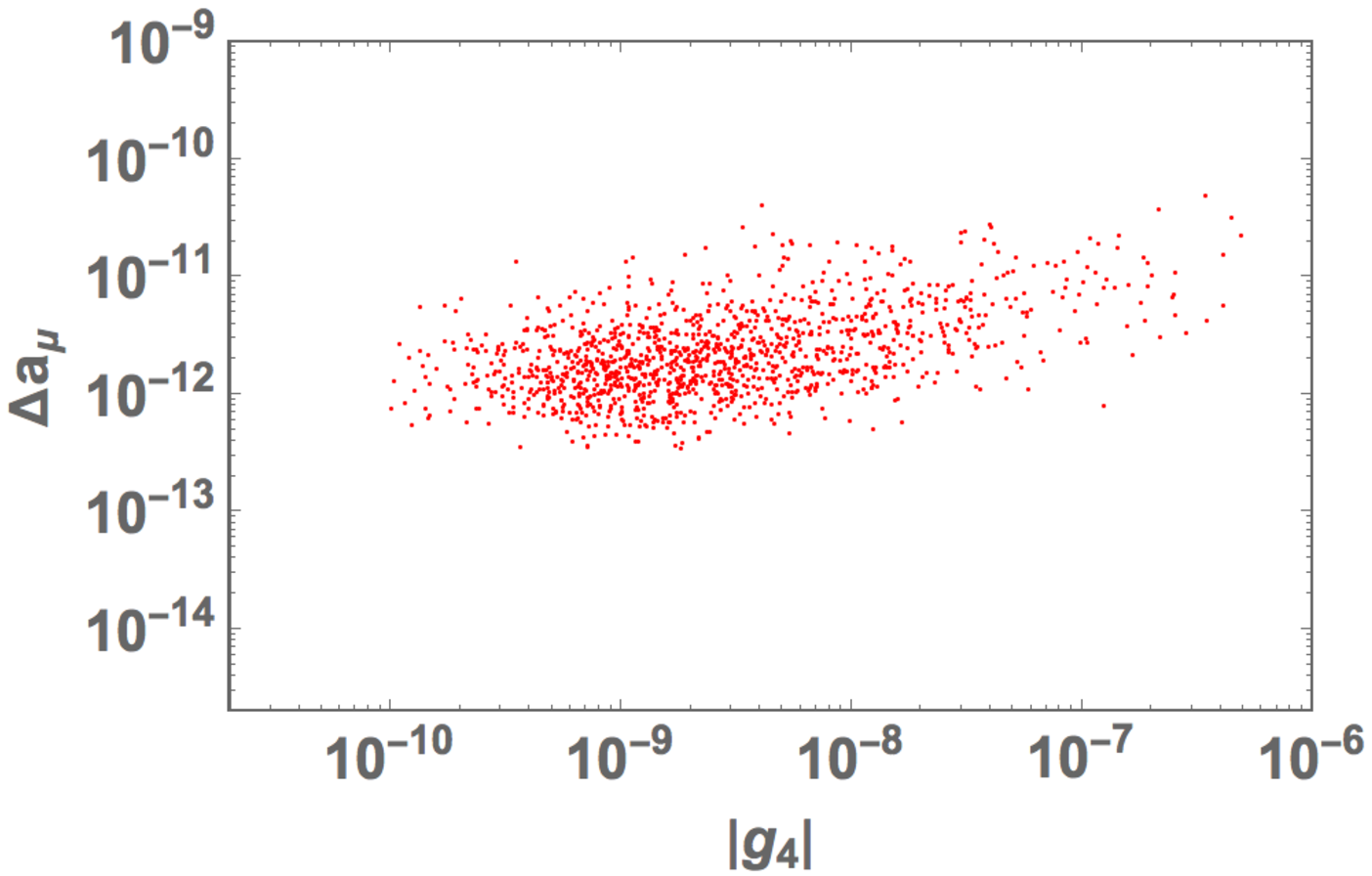} \
\includegraphics[width=80mm]{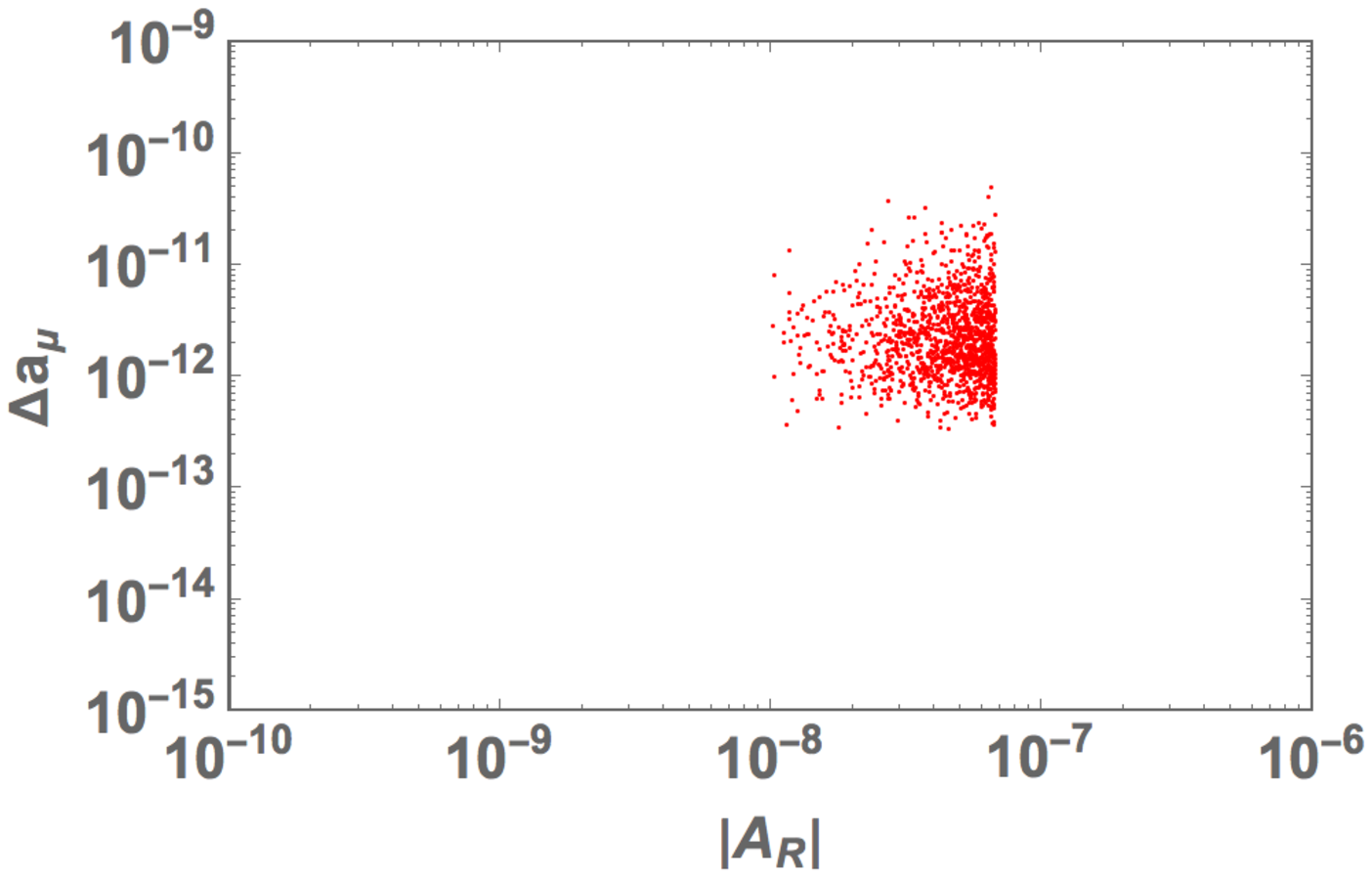}
\includegraphics[width=80mm]{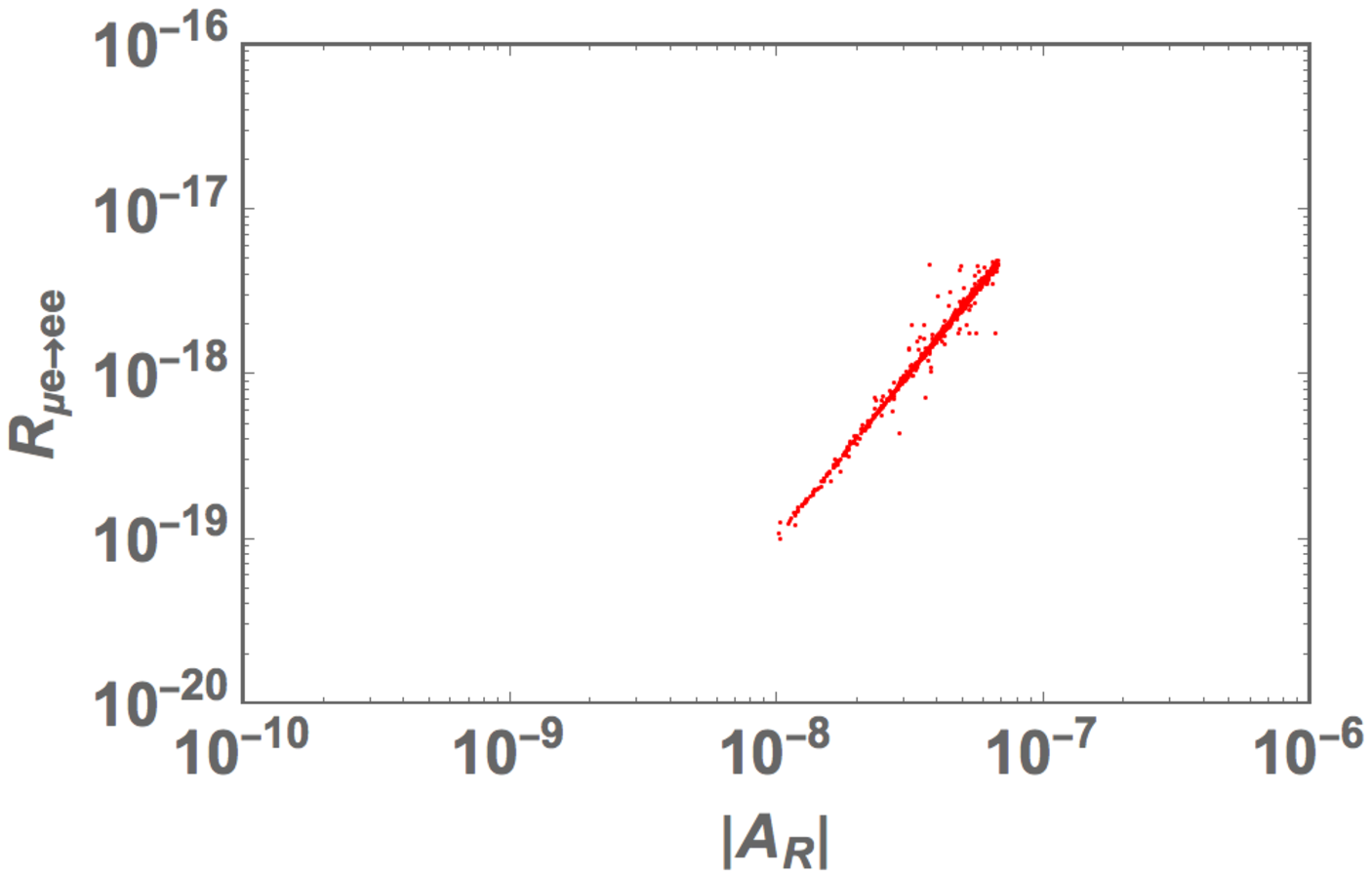} \
\includegraphics[width=80mm]{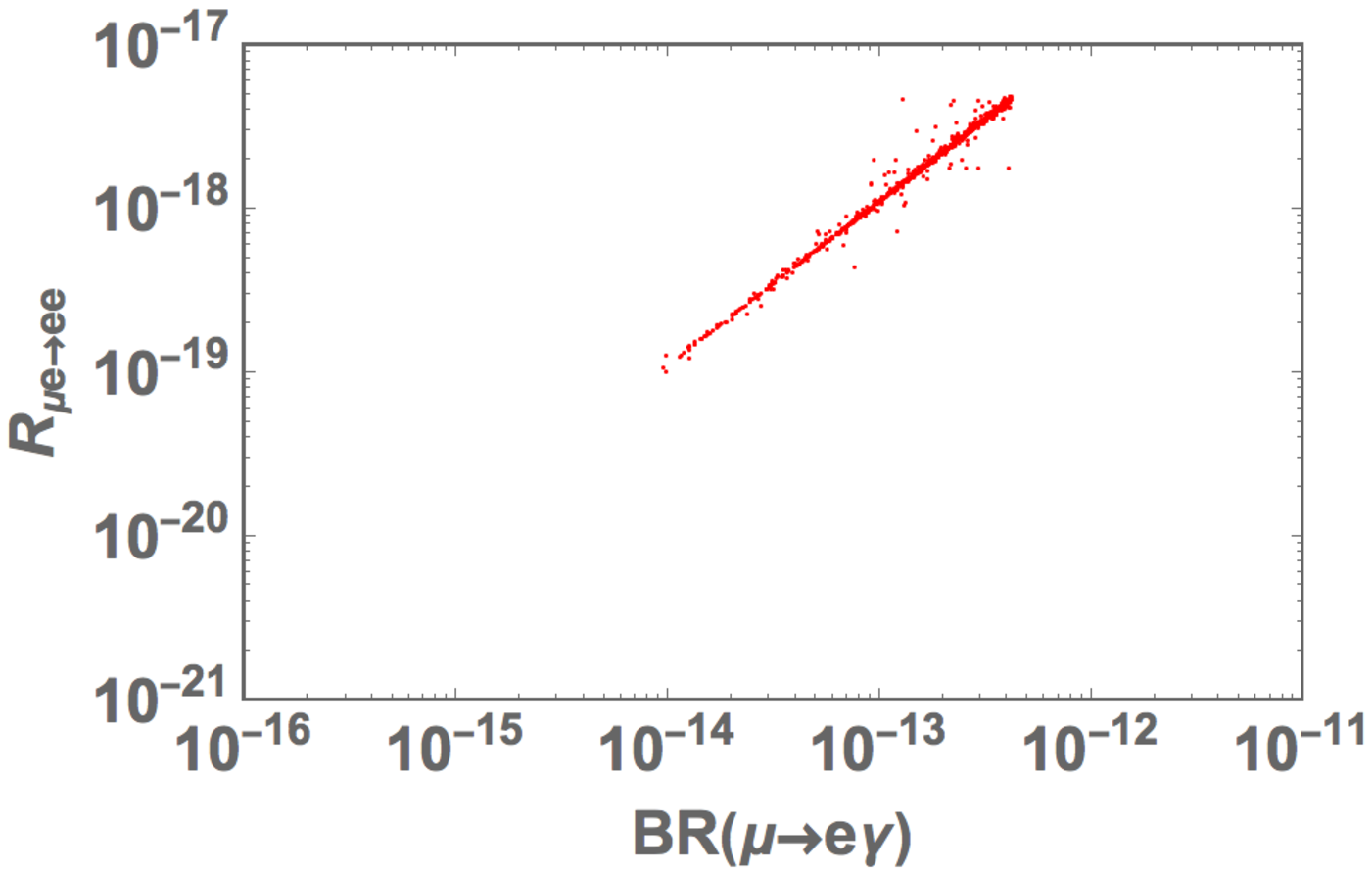}
\caption{Some correlations estimated with allowed parameter sets. Upper-left: correlation among $|g_4|$ and $\Delta a_\mu$. Upper-right: correlation among $|A_R|$ and $\Delta a_\mu$. Lower-left: correlation among $|A_R|$ and $R_{\mu e \to ee}$. Lower-right: correlation among $BR(\mu \to e \gamma)$ and $R_{\mu e \to ee}$.}
\label{fig:LFV3}
\end{figure}

In Fig.~\ref{fig:LFV1}, we provide estimated values of $BR(\ell_i \to \ell_j \gamma)$ and $\Delta a_\mu$ for allowed parameter sets 
showing correlation on $\{BR(\mu \to e \gamma), \Delta a_\mu \}$ and $\{BR(\tau \to e \gamma), BR(\tau \to \mu \gamma) \}$ plane
in left- and right-panel. We find that $\Delta a_\mu$ can be up to $\sim 4 \times 10^{-11}$ in our model. 
{In Fig.~\ref{fig:LFV2}, we also show the correlation between $BR(\mu \to e \gamma)$ and $(g_X/\Lambda_L)^2 +(g_X/\Lambda_R)^2$ on the left panel, and that between $BR(\mu \to e \gamma)$ and $BR(\mu \to e Z')$ on the right panel.
Restricted by $\mu \to e \gamma$ constraint, the maximal value of $(g_X/\Lambda_L)^2 +(g_X/\Lambda_R)^2$ is sufficiently lower than current upper bound of Eq.~\eqref{Eq: MuEX-const}, 
and the maximal value of $BR(\mu \to e Z')$ is around $10^{-13}$ that is too small to find in near-future experiments.
In the planning Mu3e experiment, the sensitivity of $BR(\mu \to e Z')$ is expected to be in the order of $10^{-8}$ \cite{Perrevoort2018a}.
Significant improvement in experimental techniques would allow us to investigate the $\mu \to e Z'$ process for our model.
In Fig.~\ref{fig:LFV3}, we show some correlations among $R_{\mu e \to e e}$, Wilson coefficients $A_R$ and $g_4$, $BR(\mu \to e \gamma)$ and $\Delta a_\mu$
estimated by using allowed parameter sets. 
In most of the parameter sets, $R_{\mu e \to e e}$ is dominantly determined by the effect of $A_R$ indicated by clear correlation in lower-left panel.
Thus it is also correlated with $BR(\mu \to e \gamma)$ as the lower-right panel since the process is induced by the operator related to $A_R$.
The effect of $g_4$ is found as deviation from the correlation.
We also find $|g_4|$ value tends to be larger when $\Delta a_\mu$ is larger as indicated by the upper-left panel.
The largest value of $R_{\mu e \to e e}$ is found to be $\sim 5 \times 10^{-18}$ where this value is the maximal value determined in almost model independently 
since the upper bound of $A_4$ is given by constraint from $\mu \to e \gamma$ process; the upper limit of $|g_4|$ is also fixed by constraint from $\mu \to eee$ process.
The expected number of stopped muons is $O\left(10^{17}\right)$ to $O\left(10^{18}\right)$ in near future experiments for $\mu^--e^-$ conversion, such as Mu2e \cite{Bartoszek2015} and COMET phase-II \cite{COMET2018}.
In these experiments, they are planning to use aluminum targets, which is less suitable for $\mu^-e^-\to e^-e^-$ due to its small proton number.
To test the value of $R_{\mu e \to ee}$,
we need next generation experiments providing larger statistics or replacement of target materials to heavier nuclei.
Notice here that the maximum $|f|$ is of the order 0.1. Thus,  it is totally safe in perturbative limit.

Before closing the section, we comment on collider bounds for new charged particle mass; $m_{E_a}$.
New charged particles $E_a$ can be produced via electroweak interactions at the LHC and they decay into charged lepton with neutral $Z_2$ odd boson $S$ as $E \to \ell S$. 
Thus our signals are similar to slepton production where slepton decays into charged lepton with neutralino. According to recent CMS analysis in ref.~\cite{CMS:2020wxd}, 
the mass upper bound is $\sim$650 GeV when neuralino is much lighter than slepton. 
On the other hand slepton can be $\sim$100 GeV when masses of slepton and neutralino are close~\footnote{For charged particles less than $\sim$100 GeV are excluded by LEP results.}. Thus our new charged particles $E_a$ have similar mass bounds.
In our analysis, we did not include collider bounds but the results for LFV processes and muon $g-2$ will not be changed significantly even if we include the bounds.

\section{Summary}

We have investigated a model based on hidden $U(1)_X$ gauge symmetry in which neutrino mass is induced at one-loop level by effects of 
interactions among particles in hidden sector and the SM leptons.
Generated neutrino masses are suppressed by loop factor and small mass difference between bosons from real and imaginary part of extra scalar field 
generated through $U(1)_X$ breaking at low scale.
Then we have formulated neutrino mass matrix, LFV processes and muon $g-2$ which are induced via interactions among SM leptons and particles in $U(1)_X$ hidden sector.

Carrying out numerical analysis, we have searched for allowed parameter sets imposing neutrino data and current LFV constraints. 
In our scenario, we can obtain sizable Yukawa couplings associated with interactions between hidden sector particles and SM leptons 
when the generated neutrino mass matrix can fit the neutrino data.
Then we have discussed LFV processes $\mu \to e Z'$, $\ell_i \to \ell_j \gamma$ and $\mu e \to e e$, and muon $g-2$ using allowed parameter sets.
It is found that these LFV processes could be tested in next generation experiments.

Before closing this paper, we would like to mention possibility of dark matter (DM) candidate.
After spontaneous symmetry breaking of $U(1)_X$, we have a remnant $Z_2$ symmetry that assures the stability of our DM candidate,
where new particles except $\varphi$ have $Z_2$ odd and the others even. As a result, there are two types of DM; $S_I$ or $\psi_1^0$.
The analysis of these DM can be found in e.g. ref.~\cite{Chiang:2017tai}.

\section*{Acknowledgments}
This research was supported by an appointment to the JRG Program at the APCTP through the Science and Technology Promotion Fund and Lottery Fund of the Korean Government. This was also supported by the Korean Local Governments - Gyeongsangbuk-do Province and Pohang City (H.O.) and the JSPS KAKENHI Grant Number JP18H01210 (Y.U.).
H. O. is sincerely grateful for the KIAS member. 
\appendix


\end{document}